\documentclass[twocolumn,pra,aps reprint, showpacs, amsmath,amssymb, aps, prl, nofootinbib]{revtex4-1}

\usepackage{amssymb}
\usepackage{amsfonts}
\usepackage{amsmath}
\usepackage{cases}
\usepackage{stmaryrd}\usepackage{tipa}
\usepackage[dvips]{graphicx}
\usepackage{dcolumn}
\usepackage{bbold}
\usepackage{graphicx}
\usepackage{subfigure}
\usepackage{dcolumn}
\usepackage{bm}
\usepackage{color}
\usepackage{xcolor}
\usepackage{CJK}
\usepackage{subfigure}
\usepackage{hyperref}
\usepackage{braket}
\usepackage{enumitem}

\def\ad{{a^\dagger}}
\newcommand{\neswarrow}{\mathrel{\text{\ooalign{$\swarrow$\cr$\nearrow$}}}}
\newcommand{\nwsearrow}{\mathrel{\text{\ooalign{$\nwarrow$\cr$\searrow$}}}}

\begin{document}

\title{Analog quantum simulation of generalized Dicke models in trapped ions}
\author{Ibai Aedo}
\author{Lucas Lamata}
\affiliation{Department of Physical Chemistry, University of the Basque Country UPV/EHU, Apartado 644, 48080 Bilbao, Spain}
\date{\today}


\begin{abstract}

We propose the analog quantum simulation of generalized Dicke models in trapped ions. By combining bicromatic laser interactions on multiple ions we can generate all regimes of light-matter coupling in these models, where here the light mode is mimicked by a motional mode. We present numerical simulations of the three-qubit Dicke model both in the weak field (WF) regime, where the Jaynes-Cummings behaviour arises, and the ultrastrong coupling (USC) regime, where rotating-wave approximation (RWA) cannot be considered. We also simulate the two-qubit biased Dicke model in the WF and USC regimes and the two-qubit anisotropic Dicke model in the USC regime and the deep-strong coupling (DSC) regime. The agreement between the mathematical models and the ion system convinces us that these quantum simulations can be implemented in the lab with current or near-future technology. This formalism establishes an avenue for the quantum simulation of many-spin Dicke models in trapped ions.

\end{abstract}


\maketitle


\section{Introduction}

Over the past few decades, quantum simulations were developed to reproduce processes of quantum systems which are difficult or even imposible to observe in the lab \cite{Georgescu_14}. The basic principle behind quantum simulations, initially introduced by Feynman \cite{Feynman_82}, is to mimic certain complex quantum dynamics using a controllable quantum system, the simulator. Many different physical platforms have been proposed for implementing quantum simulations, such as trapped ions \cite{Leibfried_03, Haffner_08, Blatt_12}, superconducting circuits \cite{Houck_12, Marcos_13, Paraoanu_14}, ultracold gases \cite{Bloch_05, Bloch_12}, quantum photonics~\cite{Lanyon_10, Guzik_12} and optical lattices \cite{Mazza_12, Szpak_12}, each of them with its own strengths and drawbacks. Here, we will consider the trapped-ion quantum technology, which is one of the most promising quantum platforms due to its high controllability and long coherence times \cite{Haffner_08}.

With the current trapped-ion technology, quantum simulations of a wide variety of models have been proposed and experimentally performed, e.g., quantum phase transitions \cite{Islam_11}, many body systems \cite{Kim_11, Casanova_12, Mezzacapo_12}, quantum field theories \cite{Casanova_11}, bosonic and fermionic interactions~\cite{Lamata_14}, relativistic quantum mechanics \cite{Lamata_07, Gerritsma_10, Casanova_10_1, Gerritsma_11, Lamata_11}, and spin models \cite{Porras_04, Friedenauer_08, Kim_10, Arrazola_16}. Following the theoretical study of the quantum Rabi model by Pedernales et al. \cite{ Pedernales_15} and its experimental realization by Lv et al. \cite{Lv_17}, here we focus on its natural generalization to multiple qubits, the so-called Dicke model. Different authors have analyzed quantum simulations of the quantum Rabi and the Dicke model in diverse quantum platforms, both theoretically \cite{Ballester_12, Mezzacapo_14, Lamata_17,Gritsev,Lesanovsky,PorrasDicke}, and experimentally \cite{Langford_17, Braumuller_17, Pietikainen_17,BollingerDicke_17}. Here, we propose analog quantum simulations of generalized Dicke models to be implemented in trapped ions as a natural extension of the quantum Rabi case~\cite{ Pedernales_15}. These models, which consist in a chain of $N$ qubits coupled to a single bosonic mode, can be reproduced with a chain of ions confined in a linear Paul trap, let us say in the $z$ direction. The ions are good approximations of two-level quantum systems and hence, they can be used as qubits. Irradiating those ions with laser beams, it is possible to generate a coupling between qubit and phonon states, such that the models of interest can be reproduced.

We present numerical simulations of the Dicke model, the biased Dicke model, and the anisotropic Dicke model, in different coupling regimes, the weak field (WF) regime and the ultrastrong coupling (USC) regime in the first two models and the USC regime and the deep-strong coupling (DSC) regime in the last case. The results show agreement between what would be expected to obtain in a physical platform and what the theoretical models predict and, therefore, we conclude that trapped ions are a flexible quantum platform to implement generalized Dicke models with current or near-future technology.


\begin{figure*}[ht]
\includegraphics[width=1\linewidth]{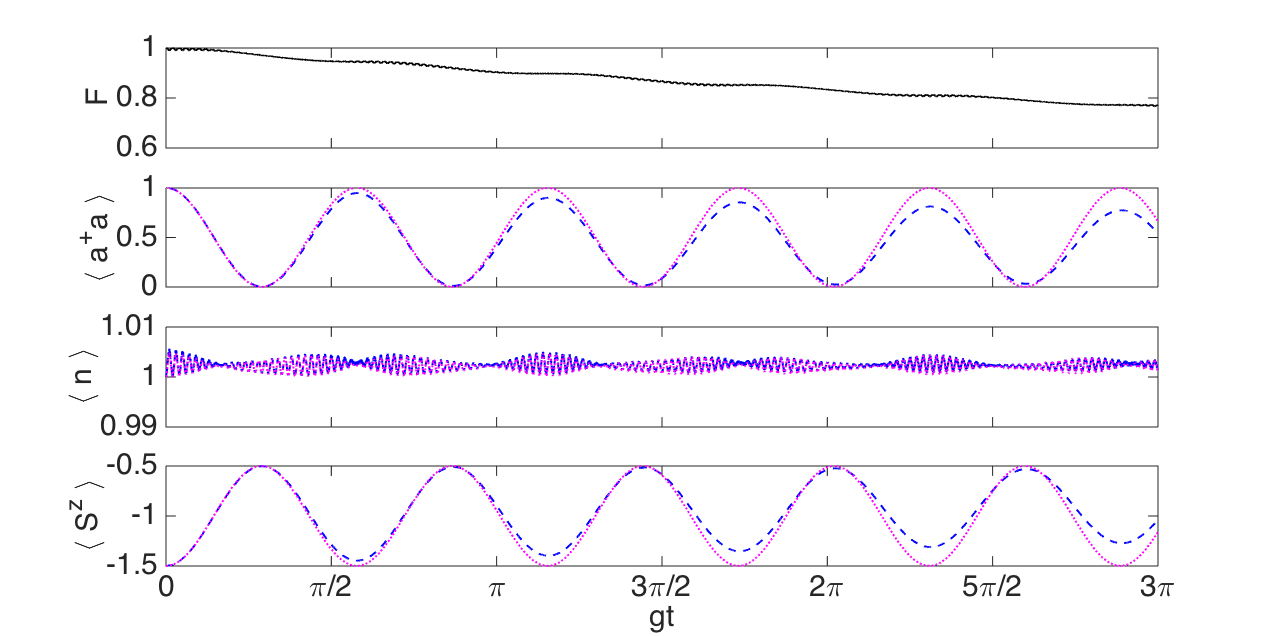}
\caption{\small{Simulation of the Dicke model with three trapped ions in the WF regime with parameters $\nu=2\pi\times3\text{MHz}$, $\omega^0=2\pi\times10^{14}\text{Hz}$, $\Omega=2\pi\times50\text{kHz}$, $\eta=0.05$, $\Gamma=\Omega\eta/100$, $\delta^r=0$ and $\delta^b=-2\pi\times125\text{kHz}$. Initial state: $\ket{1,\downarrow,\downarrow,\downarrow}$. Dotted magenta lines: Dicke model, dashed blue lines: ion system.}}
\label{Dicke3_WF}
\end{figure*}


\section{Trapped-Ion Framework}

We consider $N$ two-level ions confined in a linear trap coupled to the center of mass normal mode of a phonon field by a monochromatic laser beam. Multiple monochromatic lasers acting on the same ion can also be employed, as we will do in the following sections. The Hamiltonian of the system is ($\hbar=1$) \cite{Leibfried_03}
\begin{equation}\label{Ions_Hamiltonian}
H=\nu \ad a+\sum_{m=1}^N\left[\frac{\omega_m^0}{2}\sigma_m^z+\frac{\Omega_m}{2}\sigma_m^x\left(e^{i(k z_m-\omega_L t+\phi)}+\text{H.c.}\right)\right],
\end{equation}
where $\sigma_m^z$ and $\sigma_m^x$ are Pauli matrices associated with the internal levels of an ion, $\omega_m^0$ is the frequency of that ionic transition,  $\nu$ is the frequency of the trap, $a$ ($\ad$) is the annihilation (creation) operator of the center of mass mode, $\Omega_m$ is the Rabi frequency associated with the ion-laser coupling, and $\omega_L$, $\phi$, and $k$ are the frequency, phase, and wave number of the laser field, respectively.

Going to an interaction picture with respect to the uncoupled Hamiltonian, $\nu \ad a+\sum_{m=1}^N\omega_m^0\sigma_m^z/2$, and applying an optical rotating-wave approximation (RWA), in the so-called Lamb-Dicke regime, i.e., $\eta\sqrt{(a+\ad)^2}\ll 1$, one obtains
\begin{equation}
H^{\text{I}}=\sum_{m=1}^N\frac{\Omega_m}{2} \sigma_m^+ e^{i(\phi-\Delta_m t)}\left[1+i\eta\left(a e^{-i\nu t}+\ad e^{i\nu t}\right)\right]+\text{H.c.},
\end{equation}
where $\Delta_m=\omega_L-\omega_m^0$  is the laser detuning with respect to the internal ionic transition and $\eta=k z_0/\sqrt{N}$ is the so-called Lamb-Dicke parameter, being $z_0=\sqrt{1/2M\nu}$ the ground state width of the motional mode of a single ion of mass $M$.

In case all the ions are equal and so, $\Omega_m=\Omega$ and $\omega_m^0=\omega^0$ ($\forall m\in[1,N]$), the previous equation reduces to
\begin{equation}
H^{\text{I}}=\frac{\Omega}{2} \Sigma^+ e^{i(\phi-\Delta t)}\left[1+i\eta\left(a e^{-i\nu t}+\ad e^{i\nu t}\right)\right]+\text{H.c.},
\end{equation}
where the global operators $\Sigma^{\pm}=\sum_{m=1}^N\sigma_m^{\pm}$ have been introduced.

Choosing the laser detuning appropriately and applying a vibrational RWA, three basic resonances can be obtained. Namely, the carrier resonance ($\Delta=\delta^\text{c}$),
\begin{equation}\label{carrier}
H^{\text{c}}=\frac{\Omega^\text{c}}{2} \left(\Sigma^+ e^{i\phi^\text{c}} e^{-i\delta^\text{c} t}+\Sigma^- e^{-i\phi^\text{c}} e^{i\delta^\text{c} t}\right),
\end{equation}
the red-sideband resonance ($\Delta=-\nu+\delta^\text{r}$),
\begin{equation}\label{red_sideband}
H^{\text{r}}=i \frac{\Omega^\text{r}\eta}{2} \left( a \Sigma^+ e^{i\phi^\text{r}}e^{-i\delta^\text{r} t}- \ad \Sigma^- e^{-i\phi^\text{r}}e^{i\delta^\text{r} t}\right),
\end{equation}
and the blue-sideband resonance ($\Delta=\nu+\delta^\text{b}$),
\begin{equation}\label{blue_sideband}
H^{\text{b}}=i \frac{\Omega^\text{b}\eta}{2} \left( \ad \Sigma^+ e^{i\phi^\text{b}}e^{-i\delta^\text{b} t} - a \Sigma^- e^{-i\phi^\text{b}}e^{i\delta^\text{b} t} \right),
\end{equation}
the three of them for small values $\delta^\text{c}$, $\delta^\text{r}$ and $\delta^\text{b}$.


\section{Quantum Simulation of the Dicke Model in Trapped Ions}

We start showing how to simulate the Dicke model in a linear ion trap. The Dicke model \cite{Dicke_53, Braak_13}, which is the natural generalization of the quantum Rabi model \cite{Rabi_36, Rabi_37, Braak_11}, consists of $N$ qubits coupled to a single bosonic field mode. The contribution of the interaction between the qubits and the bosonic mode can be decomposed into a Tavis-Cummings term plus an anti-Tavis-Cummings one, giving rise to the following Hamiltonian,
\begin{align}\label{Dicke_Hamiltonian}
H_{\text{D}}=&\omega \ad a+\sum_{m=1}^N\frac{\omega_m^q}{2}\sigma_m^z+\nonumber\\
+&\overbrace{\sum_{m=1}^N g_m (a\sigma_m^++\text{H.c.})}^{\text{Tavis-Cummings}}+\overbrace{\sum_{m=1}^N g_m (\ad \sigma_m^++\text{H.c.})}^{\text{anti-Tavis-Cummings}}.
\end{align}


\begin{figure*}[ht]
\includegraphics[width=1\linewidth]{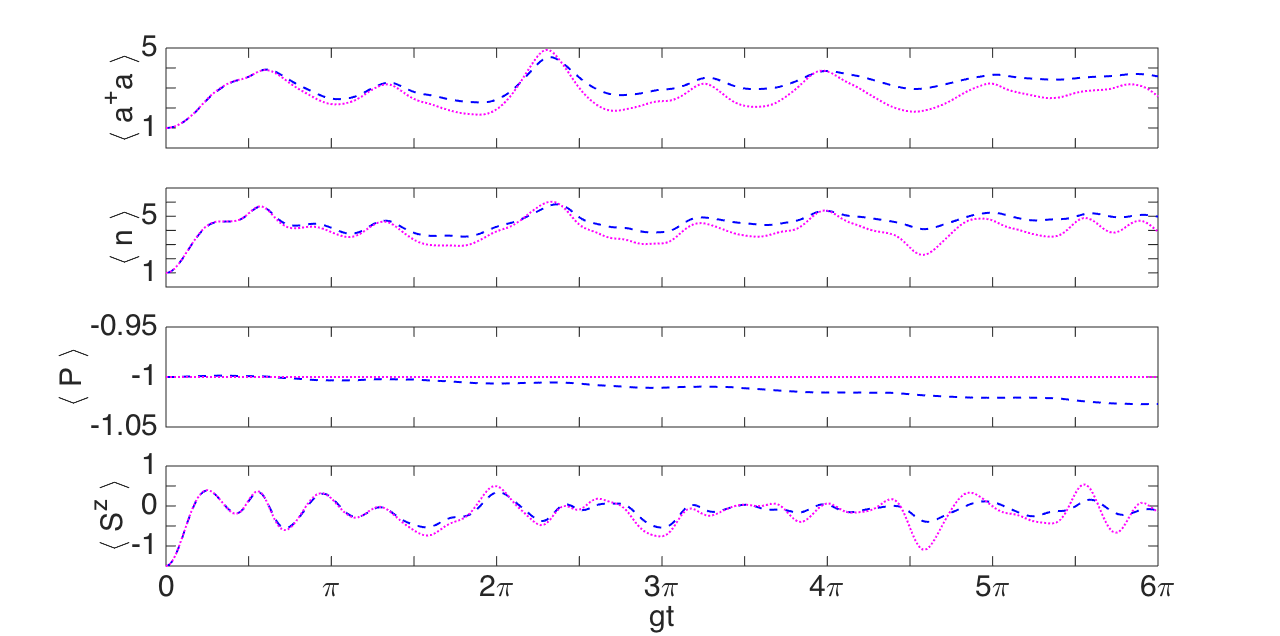}
\caption{\small{Simulation of the Dicke model with three trapped ions in the USC regime with parameters $\nu=2\pi\times3\text{MHz}$, $\omega^0=2\pi\times10^{14}\text{Hz}$, $\Omega=2\pi\times50\text{kHz}$, $\eta=0.05$, $\Gamma=\Omega\eta/100$, $\delta^r=-2\pi\times100\text{Hz}$ and $\delta^b=-2\pi\times2.7\text{kHz}$. Initial state: $\ket{1,\downarrow,\downarrow,\downarrow}$. Dotted magenta lines: Dicke model, dashed blue lines: ion system.}}
\label{Dicke3_USC}
\end{figure*}


In an interaction picture with respect to the unperturbed Hamiltonian, $\omega \ad a+\sum_{m=1}^N\omega_m^q\sigma_m^z/2$ it reads,
\begin{align}
H_{\text{D}}^{\text{I}}=&\sum_{m=1}^N g_m\left(a \sigma_m^+ e^{i(\omega_m^q-\omega)t}+\text{H.c.}\right)+\nonumber\\
+&\sum_{m=1}^Ng_m\left(\ad \sigma_m^+ e^{i(\omega_m^q+\omega)t}+\text{H.c.}\right).
\end{align}
Assuming all ionic transitions to be equal and the bosonic mode to be coupled with the same strength to every qubit, i.e., $\omega_m^q=\omega^q$ and $g_m=g$ ($\forall m\in[1,N]$), one obtains
\begin{align}\label{Dicke_interaction_picture}
H_{\text{D}}^{\text{I}}=&g\left(a \Sigma^+ e^{i(\omega^q-\omega)t}+\text{H.c.}\right)+\nonumber\\
+&g\left(\ad \Sigma^+ e^{i(\omega^q+\omega)t}+\text{H.c.}\right).
\end{align}

As Pedernales et al. pointed out in \cite{Pedernales_15} and Puebla et al. in \cite{Puebla_16} for the quantum Rabi model, i.e., the single-qubit Dicke model, the essential matter to realize a tunable quantum model is to recognize the similarity between that model and the trapped-ion system. The same is true in our multi-qubit case. The form of the Dicke model (Eq. (\ref{Dicke_interaction_picture})) is equal to the sum of the red-sideband and blue-sideband Hamiltonians of the ion system (Eqs. (\ref{red_sideband}) and (\ref{blue_sideband})), i.e., $H_{\text{D}}^{\text{I}}=H^{\text{r}}+H^{\text{b}}$, as long as the following choice is made
\begin{equation}\label{parameters_Dicke}
g=\frac{\Omega\eta}{2}\ , 
\phi^\text{r}=\phi^\text{b}=-\frac{\pi}{2}\ , 
\delta^\text{r}=\omega-\omega^q\ , 
\delta^\text{b}=-(\omega+\omega^q)\ ,
\end{equation}
where the Rabi frequencies have been renamed as $\Omega^\text{r}=\Omega^\text{b}=\Omega$. Thus, the effective frequencies of the Dicke model can be written in terms of the trapping frequency $\nu$, the frequency of the two-level systems $\omega^0$ and the laser frequencies for red and blue sidebands, $\omega^\text{r}$ and $\omega^\text{b}$,
\begin{align}
\omega&=\frac{\delta^\text{r}-\delta^\text{b}}{2}=\nu+\frac{\omega^\text{r}-\omega^\text{b}}{2}\nonumber\\
\omega^q&=-\frac{\delta^\text{r}+\delta^\text{b}}{2}=\omega^0-\frac{\omega^\text{r}+\omega^\text{b}}{2},
\end{align}
where $\omega^r=\omega^0-\nu+\delta^r$, $\omega^b=\omega^0+\nu+\delta^b$, as obtained in Ref.~\cite{Pedernales_15}.

The ion system suffers losses that have been taken into consideration in the numerical simulations. The density matrix of the ion system, $\rho$, evolves according to a master equation of the form
\begin{equation}
\frac{d\rho}{dt}=-\frac{i}{\hbar}[H,\rho]+\mathcal{L}\rho,
\end{equation}
where the Hamiltonian $H$ is given by Eq. (\ref{Ions_Hamiltonian}) and the Lindbladian operator $\mathcal{L}$ models the losses, which may be due to different decoherence sources, such as dephasing, spontaneous emission and heating. According to some trapped-ion experiments \cite{Lanyon_11}, dephasing is the dominant decoherence channel and so, a suitable representation for the Lindbladian term is \cite{Haffner_08}
\begin{equation}
\mathcal{L}\rho=\Gamma \sum_{m=1}^N\left(\sigma_m^z\rho\sigma_m^z-\rho\right).
\end{equation}


\begin{figure*}[ht]
\includegraphics[width=1\linewidth]{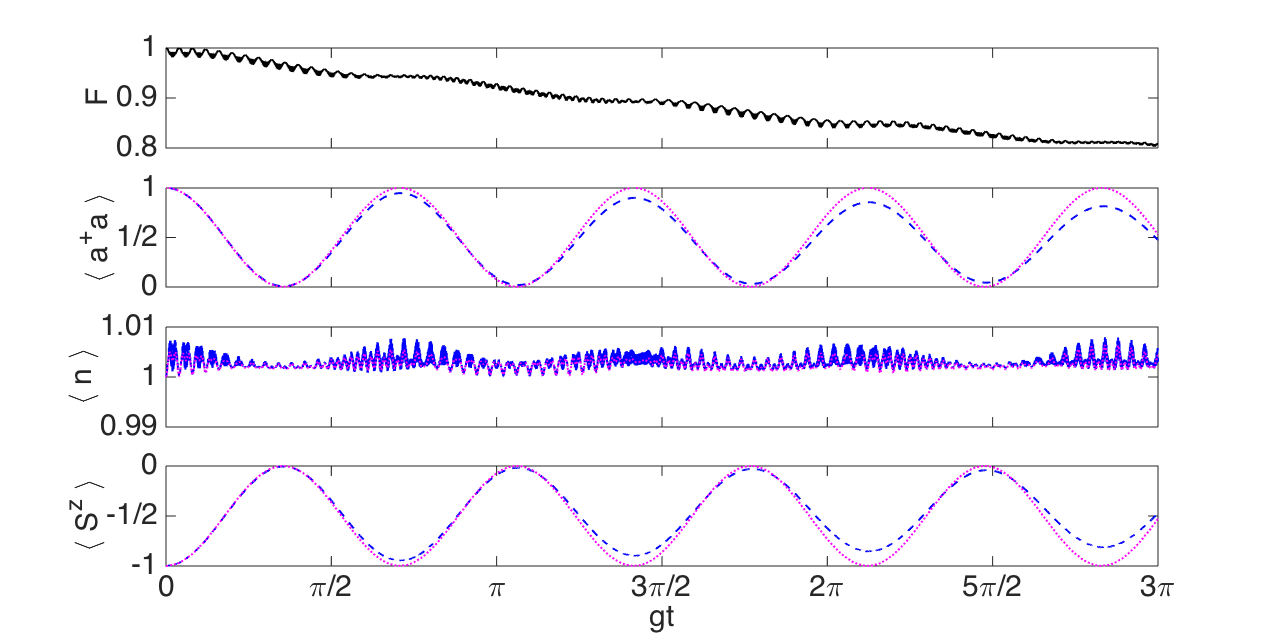}
\caption{\small{Simulation of the biased Dicke model in a two trapped-ion system in the WF regime with parameters $\nu=2\pi\times3\text{MHz}$, $\omega^0=2\pi\times10^{14}\text{Hz}$, $\Omega=2\pi\times50\text{kHz}$, $\eta=0.05$, $\Gamma=\Omega\eta/100$, $h=\Omega\eta/2$, $\delta^r=0$ and $\delta^b=-2\pi\times125\text{kHz}$. Initial state: $\ket{1,\downarrow,\downarrow}$. Dotted magenta lines: biased Dicke model, dashed blue lines: ion system.}}
\label{Biased2_WF_h_g}
\end{figure*}


Consistent with the state-of-the-art experiments with $^{40}\text{Ca}^{+}$ \cite{Gerritsma_10, Gerritsma_11}, the numerical values chosen for the trap frequency, the ionic frequency transition and the Rabi frequency are: $\nu=2\pi\times3\text{MHz}$, $\omega_0=2\pi\times10^{14}\text{Hz}$ and $\Omega=2\pi\times50\text{kHz}$. Moreover, the Lamb-Dicke parameter is in accordance with previous experiments \cite{Gerritsma_10}, $\eta=0.05$. Hence, the coupling strength of the Dicke model turns out to be $g=2\pi\times1250\text{Hz}$. The dephasing strength, $\Gamma$, which must be compared with $g$, is chosen to be $\Gamma=g/50=2\pi\times25\text{Hz}$, such that the dephasing time is $1/25 s^{-1}=40\text{ms}$, whose order of magnitude is in agreement with that of the experiments in \cite{Lanyon_11}.

Different coupling regimes can be reached just changing the laser frequencies. Two well-distinguishable regimes are studied, the WF regime and the USC regime. In the former case, the values $\delta^r=0$ and $\delta^b=-2\pi\times125\text{kHz}$ are selected, giving the effective frequencies $\omega=\omega^q=2\pi\times62.5\text{kHz}$. In the latter case, instead, the choice $\delta^r=-2\pi\times100\text{Hz}$ and $\delta^b=-2\pi\times2700\text{Hz}$ gives $\omega=2\pi\times1300\text{Hz}$ and $\omega^q=2\pi\times1400\text{Hz}$. Under the conditions $\omega\sim\omega^q$, $|\omega-\omega^q| \ll \omega,\omega^q$ and $g \ll \omega,\omega^q$, a RWA can be performed in the Dicke model, such that it resembles the Tavis-Cummings model. When $\omega=\omega^q=2\pi\times62.5\text{kHz}$, i.e., in resonance, all requirements are fulfilled such that the WF regime is achieved. It is important to remark that we are not making a RWA in the Dicke model, i.e., we are not considering the Tavis-Cummings  Hamiltonian itself. On the contrary, we are considering the full quantum Dicke Hamiltonian and making a good choice of the parameters in the ion system we are able to reach different coupling regimes, including the WF regime. Nonetheless, when $\omega=2\pi\times1300\text{Hz}$ and $\omega^q=2\pi\times1400\text{Hz}$, these frequencies are of the same order of the Rabi frequency ($g\sim\omega,\omega^q$) and the RWA cannot be considered anymore. In this case, $\omega=1.04g$ and $\omega^q=1.12g$, corresponding to the USC regime according to Pedernales et al. \cite{Pedernales_15} and Rossatto et al. \cite{Rossatto_17}.

Figures \ref{Dicke3_WF} and \ref{Dicke3_USC} show the results for the numerical simulation of the Dicke model with three ions in the WF regime and in the USC regime, respectively. The dotted magenta lines correspond to the mathematical model and the dashed blue ones to the ion system. The observables plotted are: the phonon number $\ad a$, the excitation number $n=\ad a+\sum_{m=1}^N\ket{\uparrow_m}\bra{\uparrow_m}$, the parity $P=e^{i\pi n}$, the $z$ component of the spin $S^z=\sum_{m=1}^N\sigma_m^z/2$, and the fidelity $F=\left(\text{Tr}\sqrt{\sqrt{\rho_{\text{D}}}\rho\sqrt{\rho_{\text{D}}}}\right)^2$ \cite{Jozsa_94}, a figure of merit that evaluates how similar the dynamics of the mathematical model and the dynamics of the ion system are.

\subsection{Weak Field Regime}

The results for the WF regime are in accordance with the theory of the Tavis-Cummings model. It is easy to see that for the initial state considered, $\ket{1, \sum_m^N\downarrow_m}$, the Tavis-Cummings Hamiltonian produces the single-excitation Dicke state $\ket{D_N^1}=\sum_{P}\ket{\uparrow\downarrow\dots\downarrow}/\sqrt{N}$ in the $N$-qubit case, where $P$ stands for ``all possible permutations". Thus, Rabi oscillations between $\ket{1, \sum_m^N\downarrow_m}$ and $\ket{0,D_N^1}$ arise. For $N=3$, for example, the oscillations in the spin are bounded between $-3/2$ in $\ket{1,\downarrow,\downarrow,\downarrow}$ and $-1/2$ in $\ket{0,D_3^1}$, where only one of the three qubits is up.


\begin{figure*}[ht]
\includegraphics[width=1\linewidth]{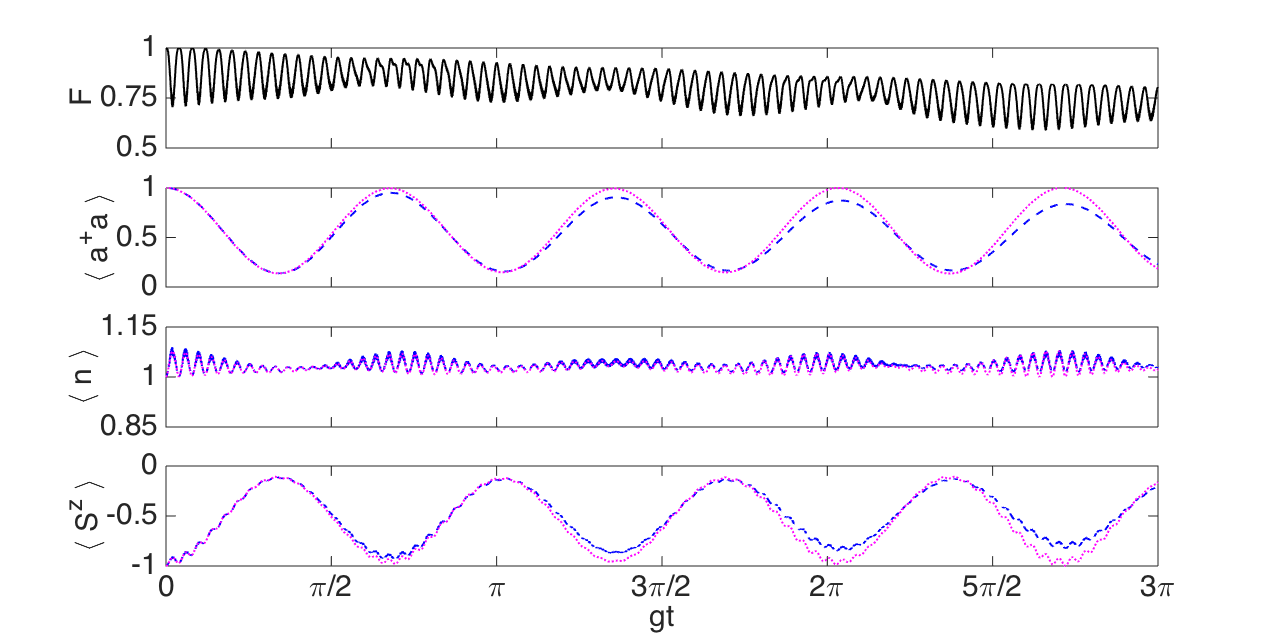}
\caption{\small{Simulation of the biased Dicke model in a two trapped-ion system in the WF regime with parameters $\nu=2\pi\times3\text{MHz}$, $\omega^0=2\pi\times10^{14}\text{Hz}$, $\Omega=2\pi\times50\text{kHz}$, $\eta=0.05$, $\Gamma=\Omega\eta/100$, $h=5\Omega\eta/2$, $\delta^r=0$ and $\delta^b=-2\pi\times125\text{kHz}$. Initial state: $\ket{1,\downarrow,\downarrow}$. Dotted magenta lines: biased Dicke model, dashed blue lines: ion system.}}
\label{Biased2_WF_h_5g}
\end{figure*}


It is well known that in the single-qubit Jaynes-Cummings model the frequency of the Rabi oscillations is $f_\text{JC}=g/\pi$. Nonetheless, in the Tavis-Cummings model it turns out to be $f_\text{TC}=\sqrt{N}f_\text{JC}$. Thus, for $N=3$ the period $T_\text{TC}$ satisfies $gT_{TC}=\pi/\sqrt{3}\approx0.58\pi$, which is in agreement with the oscillations in Fig. \ref{Dicke3_WF}. It is also worth mentioning that the excitation number, despite fluctuations, is equal to one, showing that it is a conserved observable.

The dephasing losses make the adjustment between the ions and the Dicke model worsen with the evolution of the dynamics, being the difference significant after few periods. In both the number of phonons and the spin measurement, the ions' curve is no longer fitted to the sinusoidal model. As a result of the losses, a gradual decrease of the fidelity is also perceptible.

Interestingly, the fidelity and the number of excitations show a small high frequency oscillation superposed due to a second-order effect. In resonance ($\omega=\omega^q$), the Dicke model (Eq. (\ref{Dicke_interaction_picture})) reads,
\begin{equation}
H_{\text{D}}^{\text{I}}=g\left(a \Sigma^+ +\text{H.c.}\right)+g\left(\ad \Sigma^+ e^{i2\omega t}+\text{H.c.}\right).
\end{equation}
The main term of this Hamiltonian is the first one, which corresponds to the Tavis-Cummings model and does not evolve in time. On the other hand, the second term rotates with angular frequency $2\omega$. As a consecuence, a vibrational RWA is usually considered, keeping exclusively the Tavis-Cummings contribution. However, even being the second-order contribution less important, its effect is still noticeable. The angular frequency of the superposed oscillation must be $2\omega$, which means that ($\omega/g=)50$ oscillations should be counted in an interval of length $gt=\pi$. This has been verified in Fig. \ref{Dicke3_WF}.

\subsection{Ultrastrong Coupling Regime}

In the USC regime, the Rabi oscillations disappear, that is, there are not $\ket{1, \sum_m^N\downarrow_m}\leftrightarrow \ket{0,D_N^1}$ doublets any more. It is not difficult to see that the Dicke Hamiltonian in Eq. (\ref{Dicke_Hamiltonian}) connects the initial state to a chain composed by infinitely-many states. For the single qubit case, this is the parity chain described by Casanova et al. in \cite{Casanova_10_2}. In case of multiple qubits the chain is more complex. For $N=3$, for example, it is of the form
\begin{small}
\begin{align}\label{Dicke_transitions_3qubits}
&&\ket{1,\downarrow,\downarrow,\downarrow}&&&&\ket{3,\downarrow,\downarrow,\downarrow}&&\nonumber\\
&\neswarrow&&\nwsearrow&&\neswarrow&&\nwsearrow&\nonumber\\
\ket{0,D_3^1}\hspace{2mm}&&&&\ket{2,D_3^1}\hspace{2mm}&&&&\dots\nonumber\\
&\nwsearrow&&\neswarrow&&\nwsearrow&&\neswarrow&\nonumber\\
&&\ket{1,D_3^2}\hspace{2mm}&&&&\ket{3,D_3^2}\hspace{2mm}&&\nonumber\\
&\neswarrow&&\nwsearrow&&\neswarrow&&\nwsearrow&\nonumber\\
\ket{0,\uparrow,\uparrow,\uparrow}&&&&\ket{2,\uparrow,\uparrow,\uparrow}&&&&\dots\nonumber\\
\end{align}
\end{small}
where both single-excitation and two-excitation Dicke states appear, given the last ones by $\ket{D_3^2}=(\ket{\uparrow,\uparrow,\downarrow}+\ket{\uparrow,\downarrow,\uparrow}+\ket{\downarrow,\uparrow,\uparrow})/\sqrt{3}$. As a result of the chain, and as can be seen in Fig. \ref{Dicke3_USC}, more phonons are created and, therefore, the excitation number is not conserved.


\begin{figure*}[ht]
\includegraphics[width=1\linewidth]{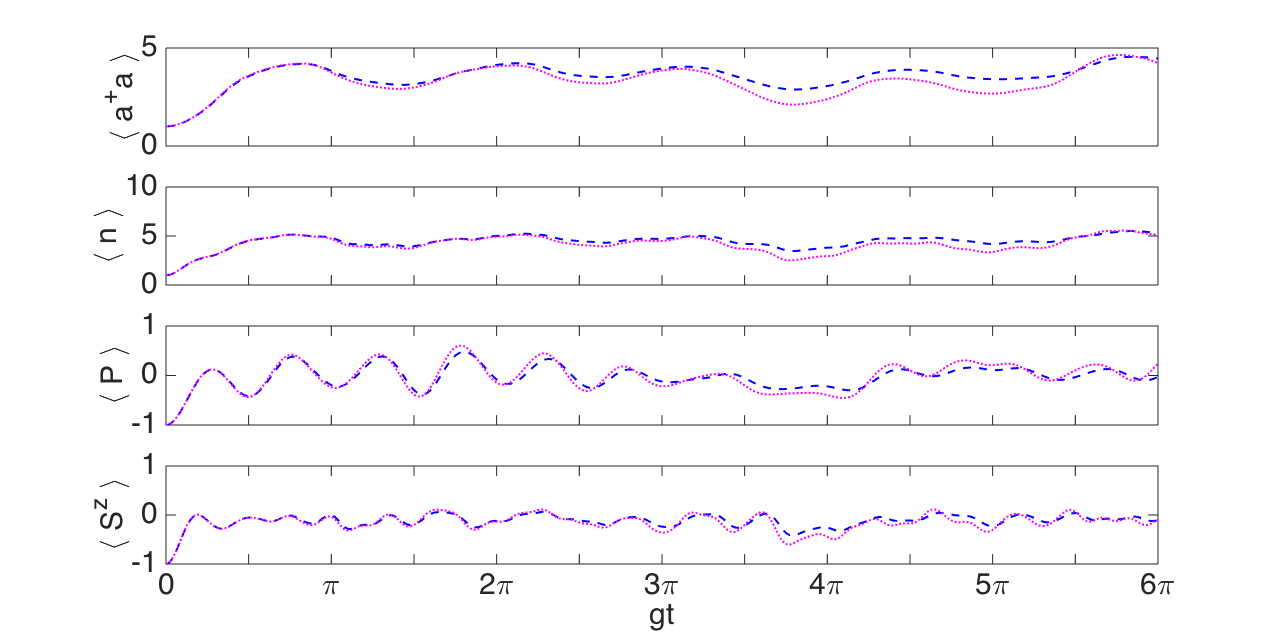}
\caption{\small{Simulation of the biased Dicke model in a two trapped-ion system in the USC regime with parameters $\nu=2\pi\times3\text{MHz}$, $\omega^0=2\pi\times10^{14}\text{Hz}$, $\Omega=2\pi\times50\text{kHz}$, $\eta=0.05$, $\Gamma=\Omega\eta/100$, $h=\Omega\eta/2$, $\delta^r=-2\pi\times100Hz$ and $\delta^b=-2\pi\times2.7\text{kHz}$. Initial state: $\ket{1,\downarrow,\downarrow}$. Dotted magenta lines: biased Dicke model, dashed blue lines: ion system.}}
\label{Biased2_USC_h_g}
\end{figure*}


The parity, however, is a conserved observable (like in the WF regime) and it is equal to $-1$, as it can be easily discerned from the initial state, which has a single excitation, $\braket{n}=1$, and hence $\braket{P}=\braket{e^{i\pi n}}=-1$. This fact is observed, up to fluctuations, in Fig. \ref{Dicke3_USC}.

Finally, it is interesting that the Rabi oscillations in the $z$ component of the spin have disappeared and collapses and revivals have emerged, as it was pointed out by Casanova et al. in \cite{Casanova_10_2} for the quantum Rabi model.


\section{Quantum Simulation of the Biased Dicke Model in Trapped Ions}

A generalization of the Dicke model is obtained considering a biased term of the form $\sum_{m=1}^Nh_m\sigma_m^x$ in Eq. (\ref{Dicke_Hamiltonian}). Assuming $\omega_m^q=\omega^q$, $g_m=g$ and $h_m=h$ ($\forall m\in[1,N]$), the following Hamiltonian arises
\begin{equation}
H_{\text{BD}}=\omega \ad a+\frac{\omega^q}{2}\Sigma^z+g(a+\ad)\Sigma^x+h\Sigma^x.
\end{equation}
The bias term becomes almost negligible when its angular frequency, $h$, is small compared to $g$. This term does not link the internal states of the qubits with the phonons, it just tries to rotate the qubits ($\ket{\downarrow}\leftrightarrow\ket{\uparrow}$). Therefore, when $h$ grows up, the bias becomes more and more important and so do the oscillations.

Moving to an interaction picture with respect to the unperturbed Hamiltonian, $\omega \ad a+\frac{\omega^q}{2}\Sigma^z$, one obtains
\begin{align}
H_{\text{BD}}^{\text{I}}&=g\left(a\Sigma^+ e^{i(\omega^q-\omega)t}+\text{H.c.}\right)+\nonumber\\
&+g\left(\ad \Sigma^+  e^{i(\omega^q+\omega)t}+\text{H.c.}\right)+\nonumber\\
&+h\left(\Sigma^+ e^{i\omega^qt}+\text{H.c.}\right).
\end{align}
This equation is of the same form of the sum of the carrier, the red-sideband and the blue-sideband Hamiltonians of the ion system (Eqs. (\ref{carrier}), (\ref{red_sideband}) and (\ref{blue_sideband})), i.e., $H_{\text{BD}}^{\text{I}}=H^\text{c}+H^\text{r}+H^\text{b}$. To perform the quantum simulation of the biased Dicke model in the trapped-ion system, it is enough to make the following choice of parameters
\begin{align}
&g=\frac{\Omega\eta}{2}\ ,h=\frac{\Omega^\text{c}}{2}\ ,\phi^\text{r}=\phi^\text{b}=-\pi/2\ ,\phi^\text{c}=0,\nonumber\\
&\delta^\text{r}=\omega-\omega^q\ ,\delta^\text{b}=-(\omega+\omega^q)\ ,\delta^\text{c}=-\omega^q,
\end{align}
with $\Omega=\Omega^\text{r}=\Omega^\text{b}$.

As it was done earlier, the choice $\delta^\text{r}=0$ and $\delta^\text{b}=-125\text{kHz}$ allows the exploration of the WF regime, while the choice $\delta^\text{r}=-2\pi\times100\text{Hz}$ and $\delta^\text{b}=-2\pi\times2700\text{Hz}$ is used to investigate the USC regime. $\delta^\text{c}$ is determined by $\delta^\text{r}$ and $\delta^\text{b}$.


\begin{figure*}[ht]
\includegraphics[width=1\linewidth]{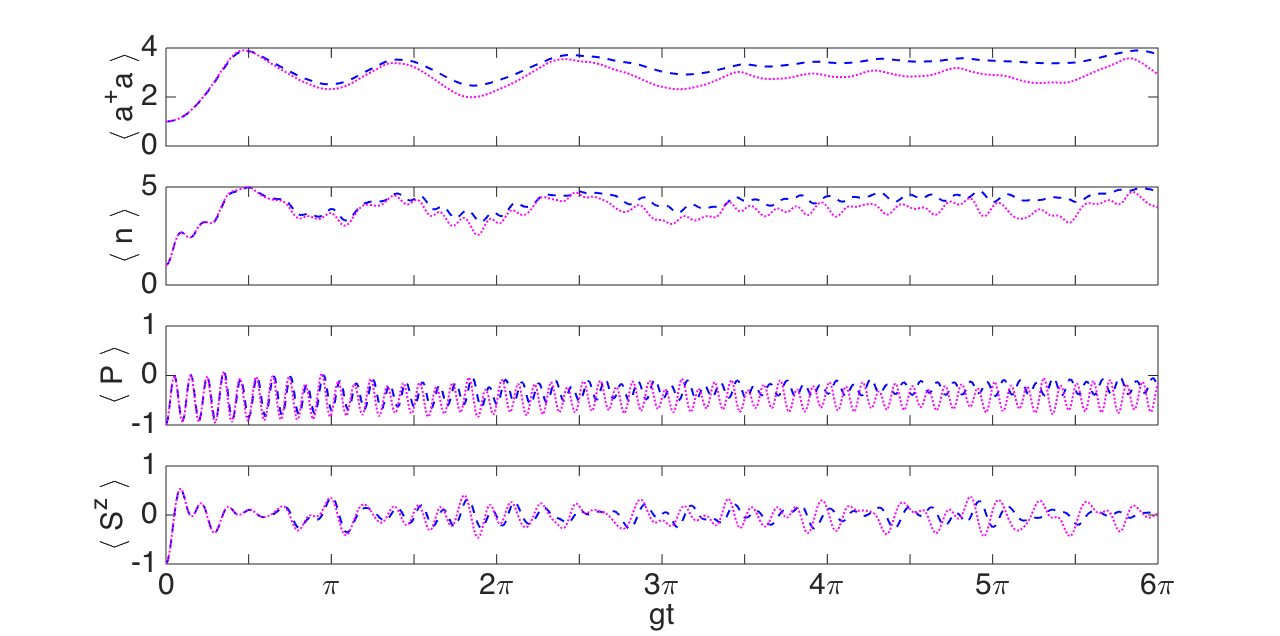}
\caption{\small{Simulation of the biased Dicke model in a two trapped-ion system in the USC regime with parameters $\nu=2\pi\times3\text{MHz}$, $\omega^0=2\pi\times10^{14}\text{Hz}$, $\Omega=2\pi\times50\text{kHz}$, $\eta=0.05$, $\Gamma=\Omega\eta/100$, $h=5\Omega\eta/2$, $\delta^r=-2\pi\times100Hz$ and $\delta^b=-2\pi\times2.7\text{kHz}$. Initial state: $\ket{1,\downarrow,\downarrow}$. Dotted magenta lines: biased Dicke model, dashed blue lines: ion system.}}
\label{Biased2_USC_h_5g}
\end{figure*}


Numerical simulations have been performed for the two-qubit biased Dicke model in the WF and USC regimes for two different strengths of the biased frequency. Figures \ref{Biased2_WF_h_g} and \ref{Biased2_WF_h_5g} show the results for the WF regime in the $h=g$ and $h=5g$ cases, respectively, while Figs. \ref{Biased2_USC_h_g} and \ref{Biased2_USC_h_5g} are the analogous for the USC regime. Again, the dotted magenta lines correspond to the mathematical model and the dashed blue ones to the ion system. 

\subsection{Weak Field Regime}

The Figs. \ref{Biased2_WF_h_g} and \ref{Biased2_WF_h_5g} of the biased Dicke model are similar to the Fig. \ref{Dicke3_WF} representing the three-ion Dicke model, but some differences are appreciated.

The second-order effect causing the superposition of high frequency oscillations in the fidelity and the excitation number is now observed with greater importance. As previously anticipated, the biased term $h \Sigma_x$ assists the rotation of the qubits and therefore, the larger the biased strength $h$ the larger and more repeated the oscillations, as it can be observed comparing Figs. \ref{Biased2_WF_h_g} and \ref{Biased2_WF_h_5g} for $h=g$ and $h=5g$, respectively.

We point out that some effects start being noticeable only when the biased frequency $h$ is strong enough and so, they are only observed in Fig. \ref{Biased2_WF_h_5g}. This is the case of the oscillations observed due to the biased term in the spin observable $S^z$.  Notice also that the mean value of the phonon number and the mean value of the spin never vanish, that is, the Rabi oscillations do not reach the $0$ value. However, they are still balanced in such a way that the excitation number remains approximately constant and equal to $1$. This can be understood as follows. When the biased term was not included, within the RWA regime the Tavis-Cummings model arised, such that, the excitation/desexcitation of the ions was accompanied by the annihilation/creation of one phonon. But now the term $h\Sigma^x=h(\Sigma^++\Sigma^-)$ is present and, therefore, the excitation/desexcitation of the spins is not necessarily accompanied by the annihilation/creation of a phonon. This last event may not happen with certain probability and consequently, the model is not fully resonant.

\subsection{Ultrastrong Coupling Regime}

In the USC regime, a similar behaviour of that observed previously in Fig. \ref{Dicke3_USC} is obtained. However, Figs. \ref{Biased2_USC_h_g} and \ref{Biased2_USC_h_5g} show that the parity is not longer a conserved operator.

Notice also that the collapses and revivals mentioned before are now better defined, and the better the greater is $h$. It is worth noting that the dynamics of the ions reproduces very well that of the model, e.g. the agreement observed in Fig. \ref{Biased2_USC_h_5g} at least up to the first revival may be enough to verify this behaviour experimentally.


\begin{figure*}[ht]
\includegraphics[width=1\linewidth]{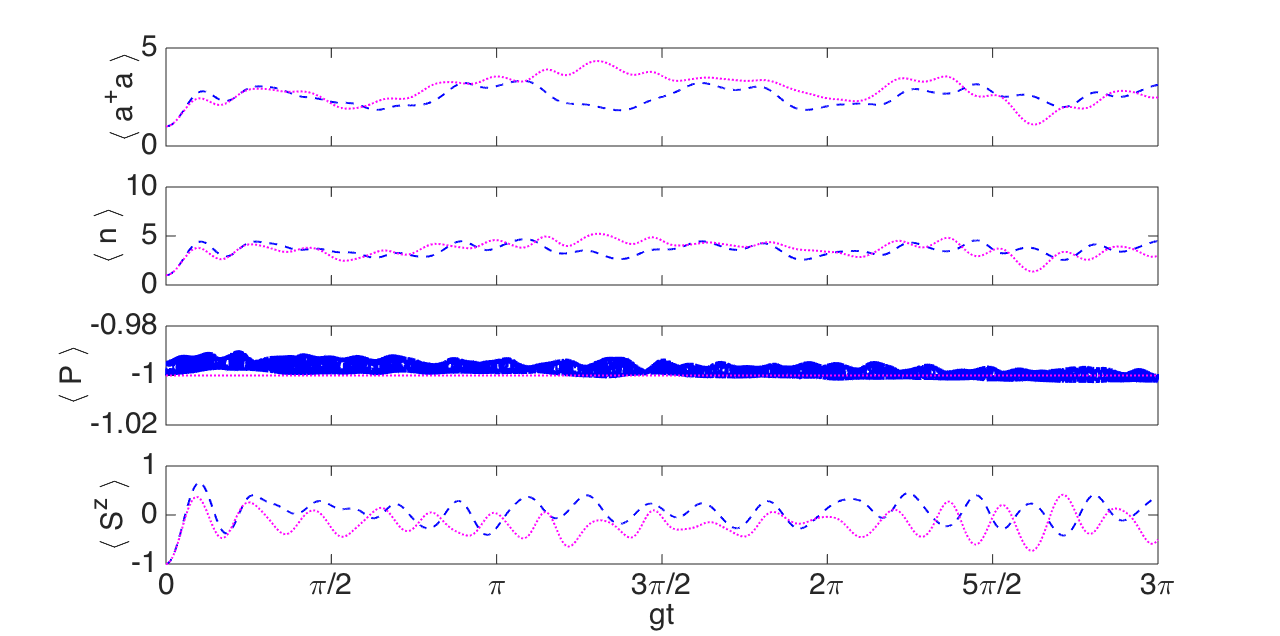}
\caption{\small{Simulation of the anisotropic Dicke model in a two trapped-ion system in the USC regime with parameters $\nu=2\pi\times3\text{MHz}$, $\omega^0=2\pi\times10^{14}\text{Hz}$, $\eta=0.05$, $\Gamma=\Omega\eta/100$, $\Omega^r=2\pi\times50\text{kHz}$, $s=3$, $\delta^r=-2\pi\times100\text{Hz}$ and $\delta^b=-2\pi\times7.7\text{kHz}$. Initial state: $\ket{1,\downarrow,\downarrow}$. Dotted magenta lines: anisotropic Dicke model, dashed blue lines: ion system.}}
\label{Anis2_USC_s3}
\end{figure*}

\begin{figure*}[t!]
\includegraphics[width=1\linewidth]{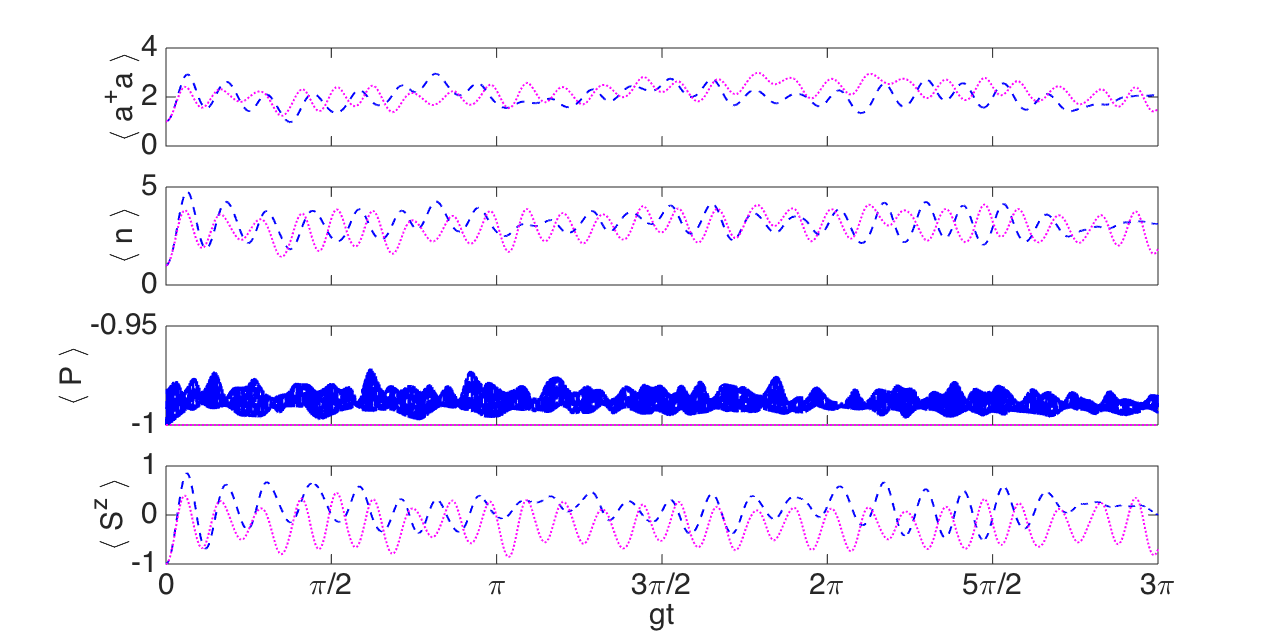}
\caption{\small{Simulation of the anisotropic Dicke model in a two trapped-ion system in the USC regime with parameters $\nu=2\pi\times3\text{MHz}$, $\omega^0=2\pi\times10^{14}\text{Hz}$, $\eta=0.05$, $\Gamma=\Omega\eta/100$, $\Omega^r=2\pi\times50\text{kHz}$, $s=5$, $\delta^r=-2\pi\times100\text{Hz}$ and $\delta^b=-2\pi\times12.7\text{kHz}$. Initial state: $\ket{1,\downarrow,\downarrow}$. Dotted magenta lines: anisotropic Dicke model, dashed blue lines: ion system.}}
\label{Anis2_USC_s5}
\end{figure*}


\begin{figure*}[ht]
\includegraphics[width=1\linewidth]{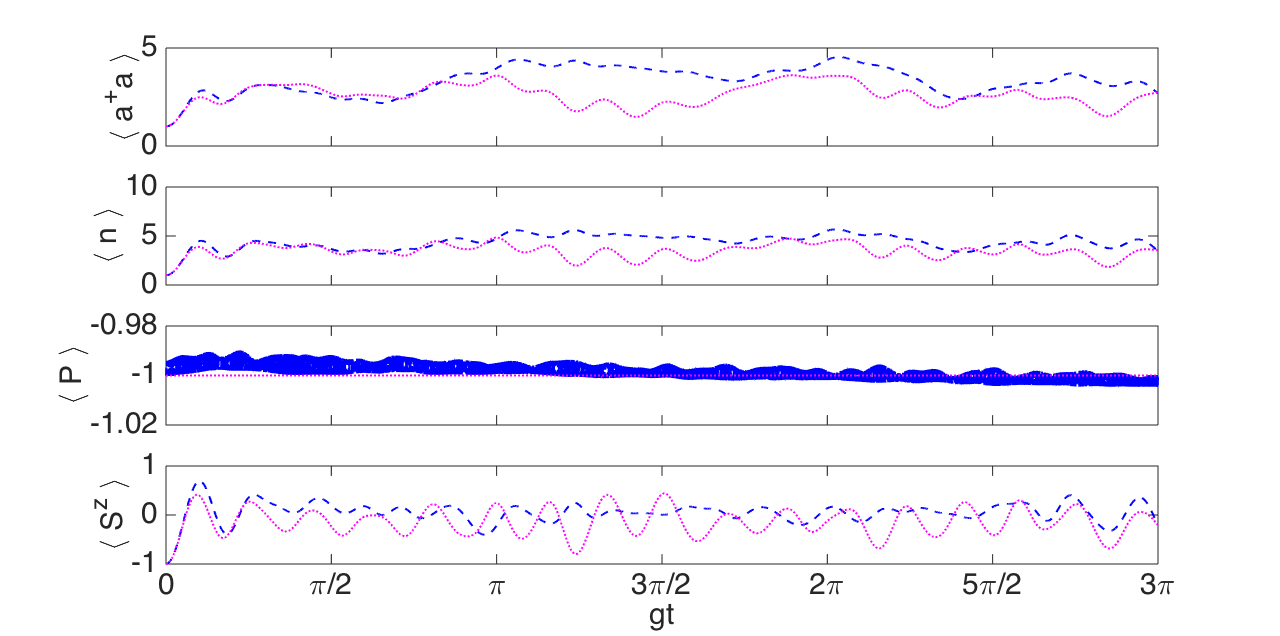}
\caption{\small{Simulation of the anisotropic Dicke model in a two trapped-ion system in the DSC regime with parameters $\nu=2\pi\times3\text{MHz}$, $\omega^0=2\pi\times10^{14}\text{Hz}$, $\eta=0.05$, $\Gamma=\Omega\eta/100$, $\Omega^r=2\pi\times50\text{kHz}$, $s=3$, $\delta^r=-2\pi\times112\text{Hz}$ and $\delta^b=-2\pi\times7238\text{Hz}$. Initial state: $\ket{1,\downarrow,\downarrow}$. Dotted magenta lines: anisotropic Dicke model, dashed blue lines: ion system.}}
\label{Anis2_DSC_s3}
\end{figure*}

\begin{figure*}[t!]
\includegraphics[width=1\linewidth]{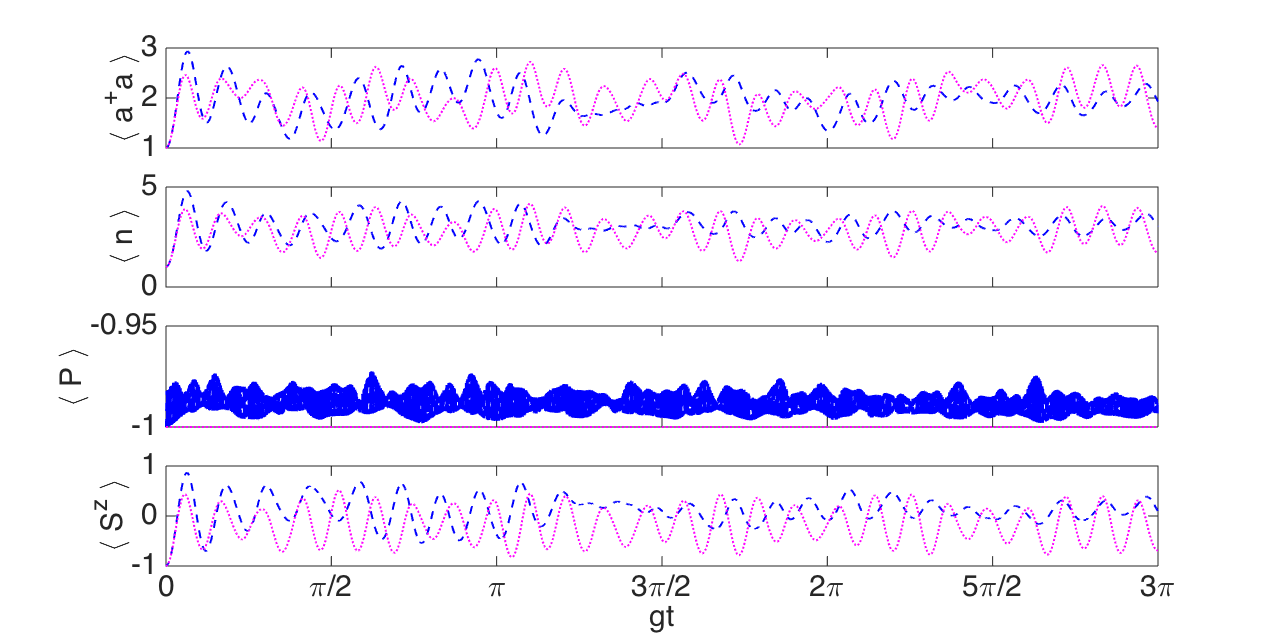}
\caption{\small{Simulation of the anisotropic Dicke model in a two trapped-ion system in the DSC regime with parameters $\nu=2\pi\times3\text{MHz}$, $\omega^0=2\pi\times10^{14}\text{Hz}$, $\eta=0.05$, $\Gamma=\Omega\eta/100$, $\Omega^r=2\pi\times50\text{kHz}$, $s=5$, $\delta^r=-2\pi\times187\text{Hz}$ and $\delta^b=-2\pi\times12063\text{Hz}$. Initial state: $\ket{1,\downarrow,\downarrow}$. Dotted magenta lines: anisotropic Dicke model, dashed blue lines: ion system.}}
\label{Anis2_DSC_s5}
\end{figure*}


\section{Quantum Simulation of the Anisotropic Dicke Model in Trapped Ions}

Following with the idea of the previous section, we now study another generalization of the Dicke model, the so-called anisotropic Dicke Model. Again, the starting point is the Dicke model (Eq. (\ref{Dicke_Hamiltonian})), but this time different weights are assigned to the the coupling strength of the Tavis-Cummings and the anti-Tavis-Cummings terms. Assuming $\omega_m^q=\omega^q$ and $g_m=g$ ($\forall m\in[1,N]$), the anisotropic Dicke model is as follows
\begin{equation}
H_{\text{AD}}=\omega \ad a+\frac{\omega^q}{2}\Sigma^z+g \left(a\Sigma^++\text{H.c.}\right)+sg\left(\ad\Sigma^++\text{H.c.}\right).
\end{equation}
When $s=0$ the Tavis-Cummings model is recovered and when $s=1$ the Dicke model is obtained. When $s\gg 1$, on the other hand, the anti-Tavis-Cummings term becomes dominant. So as to observe the influence of the anisotropy we work with $s\gg 1$, in particular, $s=3$ and $s=5$ cases are analized.

Going to an interaction picture with respect to the unperturbed Hamiltonian, $\omega \ad a+\frac{\omega^q}{2}\Sigma^z$, the following Hamiltonian arises
\begin{align}
H_{\text{AD}}^{\text{I}}=&g\left(a\Sigma^+ e^{i(\omega^q-\omega)t}+\text{H.c.}\right)+\nonumber\\
+&sg\left(\ad\Sigma^+ e^{i(\omega^q+\omega)t}+\text{H.c.}\right).
\end{align}

As in the bare Dicke model, this equation is equal to the sum of the red-sideband and the blue-sideband Hamiltonians of the ion system, i.e., $H_{\text{AD}}^{\text{I}}=H^\text{r}+H^\text{b}$, to the degree that the parameters choice of Eq. (\ref{parameters_Dicke}) is made with the following changes
\begin{equation}
g=\frac{\Omega^\text{r}\eta}{2}\ ,
\ s=\frac{\Omega^\text{b}}{\Omega^\text{r}},
\end{equation}
since $\Omega^\text{r}$ and $\Omega^\text{b}$ must differ this time.

In the present model it does not make any sense to study the WF regime because a Tavis-Cummings behaviour would be obtained and nothing about the effect of the anisotropy could be learnt. Hence, two different regimes are considered beyond the WF regime, the USC regime and the DSC regime. On the one hand, when $s=3$, choosing $\delta^r=-2\pi\times100\text{Hz}$ and $\delta^b=-2\pi\times7700\text{Hz}$ implies $\omega=1.01sg$ and $\omega_q=1.04sg$, in the USC regime; while choosing $\delta^r=-2\pi\times112\text{Hz}$ and $\delta^b=-2\pi\times7238\text{Hz}$ implies $\omega=0.95sg$ and $\omega_q=0.98sg$, in the DSC regime. On the other hand, when $s=5$, the choice $\delta^r=-2\pi\times100\text{Hz}$ and $\delta^b=-2\pi\times12700\text{Hz}$ implies $\omega=1.01sg$ and $\omega_q=1.02sg$, in the USC regime; whereas the choice $\delta^r=-2\pi\times187\text{Hz}$ and $\delta^b=-2\pi\times12063\text{Hz}$ implies  $\omega=0.95sg$ and $\omega_q=0.98sg$, in the DSC.

The simulations within the USC regime are shown in Figs. \ref{Anis2_USC_s3} and \ref{Anis2_USC_s5}, for $s=3$ and $s=5$, respectively, while the results within the DSC regime are shown in Figs. \ref{Anis2_DSC_s3} and \ref{Anis2_DSC_s5}. Although USC and DSC are different coupling regimes, in our case they are similar because $\omega=1.01sg$ is close to $\omega=0.95sg$ and $\omega_q=1.04sg$ and $\omega_q=1.02sg$ are close to $\omega_q=0.98sg$. Accordingly, for the same value of $s$, the results are much the same in both coupling regimes. Although the behaviour appears as chaotic, we can observe that the ion system reproduces the anisotropic Dicke model accurately at the very beginning of the simulation. Later, precision is lost, but at least the frequencies and the envelopes of the plots have a qualitative agreement.

\section{Experimental Considerations}

In several of the previous analyses we included numerical simulations with typical decoherence sources in trapped ions; namely, dephasing due to stray magnetic fields, as well as off-resonant excitations that may be produced by counterrotating terms present in the Hamiltonian before performing the vibrational rotating-wave approximation. However, there are other possible sources of error in trapped-ion experiments, which we estimate in this section giving evidence that they are under control.

The probability of unintended excitation of higher-frequency motional modes will be given, in the case of the nearest mode to the center of mass, i.e., the stretch mode, by $(\sqrt{N}\eta \Omega/[(\sqrt{3}-1)\nu)^2$ which is smaller than $10^{-4}$ in all cases we considered, therefore negligible.

Regarding possible collective dephasing, we point out that in some trapped-ion experiments, e.g., Monz {\it et al.,} Ref.~\cite{Monz_11}, the experimentally-observed dependence of the coherence decay is on $N^2$, instead of $N$ as corresponds to uncorrelated dephasing. Therefore, for our 2- or 3-qubit cases analyzed, the effect of collective dephasing will be a factor up to 3 times larger than the one here considered. Thus, we can safely assume that the influence of possible correlated dephasing will be of the same order of magnitude as the one in our simulations, such that it will be under control for these numbers of qubits.

With respect to heating times, we point out that in current trapped-ion experiments, as in the Innsbruck group~\cite{Blatt_12,Lanyon_11}, these can be of a few phonons per second. Given that our current proposals last in general on the order of 1 ms, we can estimate that the heating will also be negligible in this situation.


\section{Conclusions}

In recent years, the Dicke model has had renewed interest because the state-of-the-art technology makes possible its implementation in quantum simulators. In this article, we propose the analog quantum simulation of generalized Dicke models in trapped ions.

The outcomes of the Dicke model are similar to those of the Rabi model, result of being the former the straightforward generalization of the latter, but the crucial differences are the faster oscillation frequency and the generation of multipartite entanglement in the Dicke model. The plots obtained make clear that the trapped-ion platforms are flexible and fit very well to our requirements. This is revealed when the ion system reproduces the models with high accuracy in both the WF and USC regimes.

The biased Dicke model and the anisotropic Dicke model have also been simulated. The biased model exhibits not fully resonant Rabi oscillations within the WF regime and suggests collapses and revivals in the USC regime. The outcomes of the anisotropic model qualitatively reproduce the dynamics and they are sufficient to give the experimentalists an idea of what is expected to be obtained.

It should be highlighted that the time required to perform the numerical simulations grows fast with the size of the system and, consequently, ordinary computers cannot afford the numerical simulation of generalized Dicke models even for mesoscopic numbers of ions. This fact makes clear the necessity of building up quantum simulators capable of showing the behavior of many-qubit systems.


\section*{Acknowledgements}

The authors acknowledge support from MINECO/FEDER FIS2015-69983-P and Basque Government IT986-16. LL is supported by the Ram\'{o}n y Cajal Grant RYC-2012-11391.



\end{document}